\documentclass[letters,fleqn,usenatbib]{mnras}

\usepackage{newtxtext,newtxmath}
\usepackage[T1]{fontenc}

\DeclareRobustCommand{\VAN}[3]{#2}
\let\VANthebibliography\thebibliography
\def\thebibliography{\DeclareRobustCommand{\VAN}[3]{##3}\VANthebibliography}

\usepackage{graphicx}	% Including figure files
\usepackage{amsmath}	% Advanced maths commands
\usepackage{hyperref}
\usepackage{orcidlink}
\usepackage{xspace}
\usepackage{textcase}
\usepackage{xstring}
\newcommand{\citepaperI}{\protect\hyperlink{cite.Klein2024_pendulum}{Paper I}\xspace}

\newcommand{\ain}{a}
\newcommand{\out}{\rm per}
\newcommand{\aout}{a_{\out}}
\newcommand{\Pout}{P_{\rm out}}
\newcommand{\eout}{e_{\out}}
\newcommand{\mper}{m_{\rm per}}
\newcommand{\tsec}{t_{\rm sec}}
\title[EKL with Brown Hamiltonian is a simple pendulum]{Hierarchical Three-Body Problem at High Eccentricities = Simple Pendulum\\II: Octupole including Brown's Hamiltonian}

\author[Y. Y. Klein and B. Katz]{
Ygal Y. Klein\orcidlink{0009-0004-1914-5821}\thanks{E-mail: ygalklein@gmail.com (YK)}
and Boaz Katz\orcidlink{0000-0003-0584-2920}\thanks{E-mail: boaz.katz@weizmann.ac.il (BK)}
\\
Dept. of Particle Phys. \& Astrophys., Weizmann Institute of Science,
 Rehovot 76100, Israel
}

\date{Accepted XXX. Received YYY; in original form ZZZ}

\pubyear{\the\year{}}

\begin{document}
\label{firstpage}
\pagerange{\pageref{firstpage}--\pageref{lastpage}}
\maketitle

\begin{abstract}
 The very long-term evolution of the hierarchical restricted three-body problem with a massive perturber is analyzed analytically in the high eccentricity regime. Perturbations on the time scale of the outer orbit can accumulate over long timescales and be comparable to the effect of the octupole term. These perturbations are described by Brown's Hamiltonian - having different forms in the literature. We show that at the high eccentricity regime - the effect of Brown's Hamiltonian is an azimuthal precesssion of the eccentricity vector and can be solved analytically. In fact, the dynamics are equivalent to a simple pendulum model allowing an explicit flip criterion.
\end{abstract}


%%%%%%%%%%%%%%%%%%%%%%%%%%%%%%%%%%%%%%%%%%%%%%%%%%

%%%%%%%%%%%%%%%%% BODY OF PAPER %%%%%%%%%%%%%%%%%%

\section{Introduction}

In this Letter we study analytically the long-term effect of a massive perturber in a three body system with a hierarchical configuration when the inner binary is at high eccentricity. Expanding the gravitational potential up to the leading order in the semi major axis ratio (quadrupole order) - \citet{kozai1962,lidov1962} analytically solved the oscillations of eccentricity and inclinations, since then called Kozai Lidov Cycles (KLCs)\footnote{See recent historical overview including earlier relevant work by \citet{vonZeipel1910} in \citet{ito_2019}.}. The Eccentric Kozai Lidov (EKL) mechanism emerges when extending the expansion to the next order (octupole) which induces an asymmetry in the potential when the outer orbit is eccentric \citep{lithwick2011}. The octupole term allows extremely high eccentricities associated with orbital flips - change of orientation of the orbit from prograde to retrograde \citep{katz2011,naoz2011,lithwick2011,naoz2013} (for a review see \cite{naoz2016}). This effect was extensively studied numerically \citep{Ford2000,naoz2011,lithwick2011,naoz2013,li2014b,lei2022a,lei2022b} and analytically \citep{katz2011,Sidorenko2018,lei2022b,weldon2024,Klein2024_pendulum}. The EKL effect has been proposed to play a key role in a broad range of astrophysical phenomena including satellites, planets, and black hole mergers \citep{naoz2012,teyssandier2013,petrovich2015,stephan2016,liu2018,stephan21,Angelo2022,Melchor2024}. 

The common approach to study the long term evolution of a hierarchical system is the double averaging (DA) approximation.
%After expanding the gravitational potential up to some order in the small parameter of the ratio of the semi major axes an average over the inner and outer orbit is made and the evolution of the orbital parameters is governed by the averaged potential. This procedure is based on the relatively small ratio between both orbital periods and the timescale of the KLC. Analyzing the region where this assumption begins to break (for the outer orbit period) has been studied for example in \cite{luo2016} comparing the results from double averaging and full N-body simulations.
For massive perturbers, where the ratio between the period of the outer orbit to the timescale of the KLCs is comparable to, or larger than, the strength of the octupole term - the double averaging fails to reproduce the long term dynamics of the EKL \citep{luo2016,Will2021,tremaine2023}. The DA can be corrected by incorporating a correction to the Hamiltonian (Brown's Hamiltonian \citep{brown1936a,brown1936b,brown1936c,Soderhjelm1975,Cuk2004,Breiter2015,luo2016,Will2021,tremaine2023}). Recently, \citet{tremaine2023} showed that the Hamiltonians provided in \citep{brown1936c,Soderhjelm1975,Breiter2015,luo2016,Will2021} are equivalent to each other up to small oscillating perturbations due to different choices of fictitious time \citep{Will2021}. Moreover, \cite{tremaine2023} provided a new, significantly simplified, version of Brown's Hamiltonian.

% Recently, \citet{tremaine2023} analyzed the effect of non linear perturbations whose accumulation becomes important when the period of the outer orbit is not negligible with respect to the time scale of the KLCs. It was shown that an additional Hamiltonian, Brown's Hamiltonian is needed to describe these effects. Numerical experiments have reconstructed the results of N body simulations using the additional Hamiltonian. Furthermore, \citet{tremaine2023} has proved that this Hamiltonian has a gauge freedom to appear in different forms with some independently derived in numerous works in the literature \citep{brown1936a,brown1936b,brown1936c,Soderhjelm1975,Cuk2004,Breiter2015,luo2016,Will2021}.

In a previous Paper (\citet{Klein2024_pendulum}, hereafter referred to as \citepaperI), we have shown that in the high eccentricity regime, the EKL effect can be described using a simple pendulum model - with a flip criterion equivalent to the pendulum librating.

In this Letter, we extend the approach of \citepaperI using the new version of Brown's Hamiltonian provided by \cite{tremaine2023} to analytically solve the effect of Brown's Hamiltonian along with the octupole term. We show that using one of the forms of this Hamiltonian the combined effect can be described also by a simple pendulum with its velocity shifted by a constant, due to an additional azimuthal precession of the eccentricity vector.

\section{Coordinate System}

Consider a test particle orbiting a central mass $m$ on an \textit{inner} orbit with semimajor axis $a$ and eccentricity $e$ and a distant mass $m_{\text{per}}$ on an \textit{outer} orbit with $a_{\text{per}},e_{\text{per}}$ where $a/a_{\text{per}}\ll1$. Following \cite{katz2011} we align the \textit{z} axis along the direction of the total angular momentum vector (which is the angular momentum vector of the outer orbit since the inner orbit is of a test particle). The \textit{x} axis is directed towards the pericenter of the (constant) outer orbit.
The dynamics of the test particle can be parameterized
by two dimensionless orthogonal vectors $\mathbf{j}=\mathbf{J}/\sqrt{GMa}$, where $\mathbf{J}$ is the specific angular momentum vector, and
$\mathbf{e}$ a vector pointing in the direction of the pericenter
with magnitude $e$. Using the variables $i_e$ and $\Omega_e$ as in \cite{katz2011} the eccentricity vector $\mathbf{e}$ is given by 
\begin{equation}
  \mathbf{e}=e\left(\sin i_e \cos \Omega_e,\sin i_e \sin\Omega_e,\cos i_e\right).
\end{equation}
%Note that it is usually written using the argument of pericenter $\omega$ and longitude of ascending node $\Omega$ as %$e\left(\cos\omega\cos\Omega-\sin\omega\sin\Omega\cos i,\cos\omega\sin\Omega+\sin\omega\cos\Omega\cos i,\sin\omega\sin i\right)$.
% \begin{equation*}
% \mathbf{e}=e\left(\begin{array}{c}
% \cos\omega\cos\Omega-\sin\omega\sin\Omega\cos i\\
% \cos\omega\sin\Omega+\sin\omega\cos\Omega\cos i\\
% \sin\omega\sin i
% \end{array}\right)  
% \end{equation*}

Expanding the perturbing potential to the third order term in the small parameter $a/a_{\text{per}}$ (the octupole term), averaging over the inner orbit, calculating the oscillations of $\mathbf{j}$ and $\mathbf{e}$ during an outer orbit and then averaging over the outer orbit results with the following potential:
% \begin{equation}
% \Phi_{{\text{per}}}=\Phi_{0}\left(\phi_{\text{quad}}+\epsilon_{\text{oct}}\phi_{\text{oct}}+\epsilon_{\text{SA}}\phi_{CDA}\right)\label{eq:phi_Per}
% \end{equation}
\begin{equation}
\Phi_{{\text{per}}}=\Phi_{0}\left(\phi_{\text{quad}}+\epsilon_{\text{oct}}\phi_{\text{oct}}+\epsilon_{\text{SA}}\phi_{\text{HB3}}\right)\label{eq:phi_Per}
\end{equation}
where (e.g, \citet{katz2011,liu2014,petrovich2015,tremaine2023,luo2016})
\begin{equation}
 \phi_{\text{quad}}=\frac{3}{4}\left(\frac{1}{2}j_{z}^{2}+e^{2}-\frac{5}{2}e_{z}^{2}-\frac{1}{6}\right),\label{eq:phi_quad}
\end{equation}
\begin{equation}
 \phi_{\text{oct}}=\frac{75}{64}\left(e_{x}\left(\frac{1}{5}-\frac{8}{5}e^{2}+7e_{z}^{2}-j_{z}^{2}\right)-2e_{z}j_{x}j_{z}\right),\label{eq:phi_oct}
\end{equation}
% \begin{align}\label{eq:CDAPotential}
% \phi_{\text{CDA}}=-&\Big(\frac{27}{64}j_z[(1-j_z^2)/3+8e^2-5e_z^2]+\cr
% 		&+\frac{3\eout^2}{64}[e_z(10j_xe_x-50j_ye_y)+\cr&j_z(5j_x^2-j_y^2+65e_x^2+35e_y^2)]\Big),
% \end{align}
and (Equation 67 of \citet{tremaine2023})
\begin{equation}\label{eq:HB3Potential}
\phi_{\text{HB3}}=-\frac{27}{64}\Big(1+\frac{2}{3}\eout^2\Big)j_z[(1-j_z^2)/3+8e^2-5e_z^2],
\end{equation}

% and the constant parameters
\begin{equation}
 \Phi_{0}=\frac{Gm_{\text{per}}a^{2}}{a_{\text{per}}^{3}\left(1-e_{\text{per}}^{2}\right)^{\frac{3}{2}}} \label{eq:phi0}
\end{equation}
\begin{equation}
 \epsilon_{\text{oct}}=\frac{a}{a_{\text{per}}}\frac{e_{\text{per}}}{1-e_{\text{per}}^{2}} \label{eq:epsilon_oct}
\end{equation}
\begin{equation}
 \epsilon_{\text{SA}}=(\frac{\ain}{\aout})^{3/2}\frac{1}{(1-\eout^2)^{3/2}}\frac{\mper}{[(m+\mper)m]^{1/2}} \label{eq:epsilon_SA}
\end{equation}
with $\Phi_{0}$, $\epsilon_{\text{oct}}$ and $\epsilon_{\text{SA}}$ constant.
Time and its derivatives are expressed in units of the secular timescale
\begin{equation}
 t_{\rm sec}=\sqrt{GMa}/\Phi_0 \label{eq:tsec}
\end{equation}
using $\tau\equiv t/t_{sec}$ and obtaining $\epsilon_{\text{SA}}=\frac{\Pout}{2\pi \tsec}$ measuring the ratio between the period of the outer orbit and the secular Kozai Lidov timescale.

\section{Slow variables}
When $\epsilon_\text{oct}=0$ (the periodic analytically solved KLCs) the perturbing potential is axisymmetric (with respect to the \textit{z} axis) admitting a constant of the motion, $j_z$, which limits the eccentricity through the constraint $j>j_z$. When neglecting $\epsilon_{\text{SA}}$ KLCs are classified by the values of the constants $j_z$ and
\begin{align}
C_{K}&=\frac{4}{3}\phi_{\text{quad}} + \frac{1}{6} - \frac{1}{2}j^2_z\cr
&=e^{2}-\frac{5}{2}e^2_z. \label{eq:CK}
\end{align}
When $C_K<0$, the argument of pericenter of the inner orbit, $\omega$, librates around $\frac{\pi}{2}$ or $-\frac{\pi}{2}$ (\textit{librating} cycles), and when $C_K>0$, it goes through all values $\left[0,2\pi\right]$ (\textit{rotating} cycles).

In the problem we study here with $\epsilon_\text{oct}>0$ and without neglecting $\epsilon_{\text{SA}}$, $C_K$ and $j_z$ are no longer constant. The three slow variables $j_z,C_K$ and $\Omega_e$ change slowly on a timescale of $\sim t_{\text{sec}}/\sqrt{\epsilon_\text{oct}}$ (see Eqs. \ref{eq:epsilon_oct}-\ref{eq:tsec}) \citep{antognini2015}. As in \cite{katz2011} we focus on the regime of high eccentricity, i.e $\left|j_z\right|\ll1$.

Up to leading order in $j_z$, we have from Equation \ref{eq:phi_Per} and Equation 20 of \citet{tremaine2023}
\begin{equation}
\begin{aligned}
  \dot{\Omega}_{e} & = f_{\Omega} j_{z}-\epsilon_{\text{SA}}\left(1+\frac{2}{3}e_{{\text{per}}}^{2}\right)f_{\text{SA}}
  \\
 \dot{j}_z & = -\epsilon_\text{oct} f_j \sin \Omega_e
\end{aligned}
\end{equation}
where
\begin{equation}
  f_{\text{SA}}=\frac{27}{64}\left(\frac{1}{3}+8C_{K}+15e_{z}^{2}\right)
\end{equation}
and \citep{katz2011}
$f_j=\left(75/64\right)e\sin i_e\left(\frac{1}{5}-e^2\left(\frac{8}{5}-7\cos ^2 i_e\right)\right)$,\\ $f_{\Omega}=3\left(4-3/\sin^2i_e\right)/4$.

\section{Averaged Equations}\label{sec:averaged_equations}
Averaging over rotating KLCs at $j_z=0$ (up to the leading order in $j_z$ and $\epsilon_\text{oct}$) (librating KLCs with $\left|j_z\right|\ll1$ accumulate change in $j_z$ on a much longer timescale \citep{katz2011}), the averaged equations for the long term evolution of $\Omega_e$ and $j_z$ are 
\begin{equation}
\begin{aligned}
 \dot{\Omega}_e & = \left\langle f_\Omega \right\rangle j_z-\epsilon_{\text{SA}}\left(1+\frac{2}{3}e_{{\text{per}}}^{2}\right)\left\langle f_{\text{SA}}\right\rangle\\
 \dot{j}_z & = -\epsilon_\text{oct} \left<f_j\right> \sin \Omega_e
\end{aligned}
\label{eq:averaged_equations}
\end{equation}
where
\begin{equation}
\begin{aligned}
 \left\langle f_{\text{SA}}\right\rangle & =\frac{27}{64}\left(\frac{1}{3}+8C_{K}+15\left\langle e_{z}^{2}\right\rangle \right), \\
 \left\langle e_{z}^{2}\right\rangle & =\frac{2}{5}\left(1-C_{K}\right)\frac{E\left(x\right)-\left(1-x\right)K\left(x\right)}{xK\left(x\right)}
\end{aligned}\label{eq:averaged_functions_ours}
\end{equation}
and \citep{katz2011}
\begin{equation}
\begin{aligned}
 \left\langle f_{\Omega}\right\rangle & =\frac{6E\left(x\right)-3K\left(x\right)}{4K\left(x\right)}, \\
 \left\langle f_{j}\right\rangle & =\frac{15\pi}{128\sqrt{10}}\frac{1}{K\left(x\right)}\left(4-11C_{K}\right)\sqrt{6+4C_{K}},\\
 x & =3\frac{1-C_{K}}{3+2C_{K}}
\end{aligned}\label{eq:averaged_functions_Katz2011}
\end{equation}
and $K\left(m\right)$ and $E\left(m\right)$ are the complete elliptic functions of the first and second kind, respectively.

\section{Simple Pendulum}\label{sec:simple_pendulum}

As presented in \citepaperI, since $\Phi_{{\text{per}}}$ is constant and $\epsilon_{\text{oct}}\phi_{\text{oct}}$, $\epsilon_{\text{SA}}\phi_{\text{HB3}}$ and $j_z$ are small - $C_K$ and therefore $\left\langle f_{\Omega}\right\rangle,\left\langle f_{j}\right\rangle,\left\langle f_{SA}\right\rangle$ are approximately constant, Equations \ref{eq:averaged_equations} are equations of a simple pendulum with angle $\phi=\Omega_e$ where the product $\left\langle f_{\Omega}\right\rangle\left\langle f_{j}\right\rangle>0$ and $\phi=\Omega_e+\pi$ otherwise. The velocity of the pendulum is given by 
\begin{equation}
  \dot{\phi} = \left\langle f_\Omega \right\rangle j_z-\Delta\label{eq:phidot}
\end{equation}
with 
\begin{equation}
  \Delta=\epsilon_{\text{SA}}\left(1+\frac{2}{3}e_{{\text{per}}}^{2}\right)\left\langle f_{\text{SA}}\right\rangle\label{eq:Delta}
\end{equation}
a positive constant. The pendulum has a constant energy 
\begin{equation}
  E=\frac{1}{2} \left(\dot{\phi}^0\right)^2 + V^0\label{eq:pendulum_energy}
\end{equation}
where the potential $V$ is given by 
\begin{equation}
  V=\epsilon_{\text{oct}}\left|\left\langle f_\Omega\right\rangle\left\langle f_{j}\right\rangle\right|\left(1-\cos\phi\right)\label{eq:pendulum_potential}.
\end{equation}

\begin{figure}
 \begin{centering}
 \includegraphics[scale=0.22]{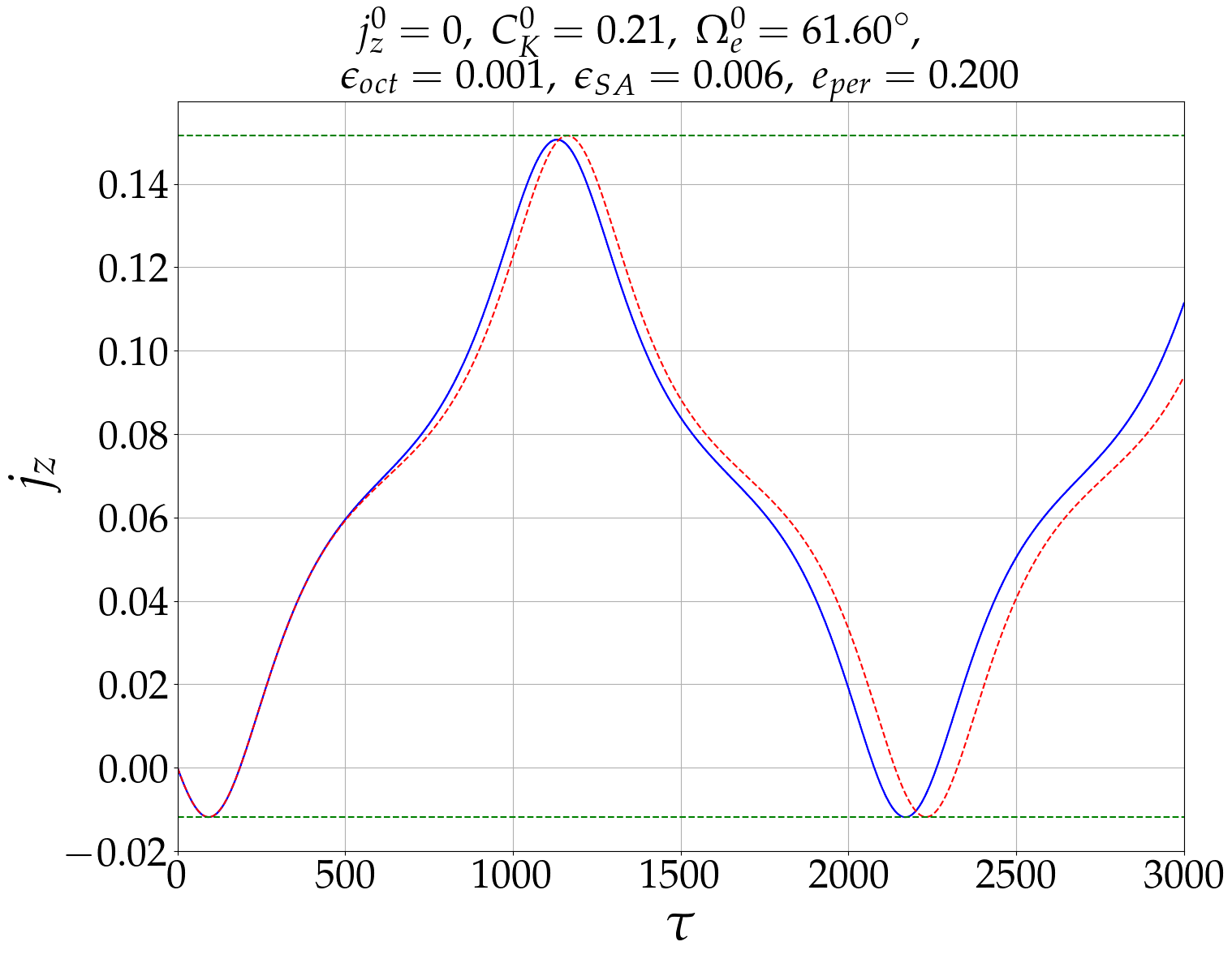}
 \par\end{centering}
 % \begin{centering}
 % \includegraphics[scale=0.25]{epsilonoct_0.001_jz_vs_Omega_e_up_to_tau_3161.84_no_box.png}
 % \par\end{centering}
 \caption{Results of a numeric integration of the double averaged equations (blue) along with the result of a simple pendulum (Equations \ref{eq:averaged_equations}, red). The values of the initial conditions and $\epsilon_{\text{oct}},\epsilon_{\text{SA}}$ and $e_{{\text{per}}}$ are shown above the plot. Shown is $j_z$ as a function of (normalized) time. The two green horizontal lines are the maximum and minimum values of $j_z$ of the simple pendulum.
 \label{fig:jz_vs_tau_example}}
\end{figure}

An example of a numerical integration of the full double averaged equations (Equations 20 in \cite{tremaine2023}) compared with the solution of Equations \ref{eq:averaged_equations} with the approximation of constants $\left\langle f_{\Omega}\right\rangle,\left\langle f_{j}\right\rangle,\left\langle f_{\text{SA}}\right\rangle$ (evaluated at $C^0_K$) is shown in Figure \ref{fig:jz_vs_tau_example}. As can be seen - for the example shown - the long term evolution of $j_z$ is successfully approximated by equations of a simple pendulum.

\subsection{Scaling}\label{subsec:scaling}
The potential for Brown Hamiltonian in its HB3 form, Equation \ref{eq:HB3Potential}, is axis symmetric (with respect to the $\textit{z}$-axis). Denoting the longitude of ascending node as $\Omega$ leads, at low $j_z$, to $\dot{\Omega}=\dot{\Omega}_e$ being independent of $\Omega_e$ making the effective equations for the slow variables to be equations of a simple pendulum with the velocity shifted by a constant ($\Delta$, Equation \ref{eq:Delta}). As presented in \citepaperI, the velocity in the pendulum model, i.e $j_z$ there - scales as $\sqrt{\epsilon_\text{oct}}$. Therefore, adding the constant $\Delta$ to the effective equations to account for Brown's Hamiltonian - makes the equations scale with a scaling factor of 
\begin{equation}
  s=\frac{\epsilon_\text{SA}\left(1+\frac{2}{3}e^2_{\text{per}}\right)}{\sqrt{\epsilon_\text{oct}}}\label{eq:scaling_factor}.
\end{equation}

\section{flips}\label{sec:flips}
A flip - zero crossing of $j_z$ - happens when the product $j^{\text{max}}_zj^{\text{min}}_z < 0$. Using Equation \ref{eq:phidot} we have for example where $\left\langle f_\Omega \right\rangle > 0$, i.e for $C^0_K \gtrsim 0.112$
\begin{equation}
  j^{\text{max}}_z = \frac{1}{\left\langle f_\Omega \right\rangle}\left(\dot{\phi}^{\text{max}} +\Delta \right)\label{eq:jzmax}
\end{equation}
and similar expression for $j^{\text{min}}_z$ (and vice-versa when $\left\langle f_\Omega \right\rangle < 0$ at $C^0_K \lesssim 0.112$).
Since the analytic model is a simple pendulum model, obtaining $\dot{\phi}^{\text{max}}$ and $\dot{\phi}^{\text{min}}$ is straightforward. Given $\epsilon_{\text{oct}},\epsilon_{\text{SA}}$ and $e_{\text{per}}$ and initial values $j^0_z,C^0_K$ and $\Omega^0_e$, the simple pendulum can be in either mode, {\em{Libration}} when $E < 2 \epsilon_{\text{oct}} \left| \left\langle f_{j} \right\rangle \left\langle f_{\Omega} \right\rangle \right|$ or otherwise in {\em{Rotation}}.
% \subsection{Maximal Deviation of $\MakeLowercase{j}_\MakeLowercase{z}$ from $\MakeLowercase{j}^0_\MakeLowercase{z}=0$}\label{subsec:jzmax_from_jz0}
\subsection{Maximal Deviation of $j_z$ from $j^0_z=0$}\label{subsec:jzmax_from_jz0}

% \subsection{Maximal Deviation of $\textlowercase{j}_\textlowercase{z}$ from $\textlowercase{j}^0_\textlowercase{z}=0$}\label{subsec:jzmax_from_jz0}

% \subsection{Maximal Deviation of \textlowercase{$j_z$} from \textlowercase{$j^0_z=0$}}\label{subsec:jzmax_from_jz0}

\begin{figure}
 \begin{centering}
 \includegraphics[scale=0.45]{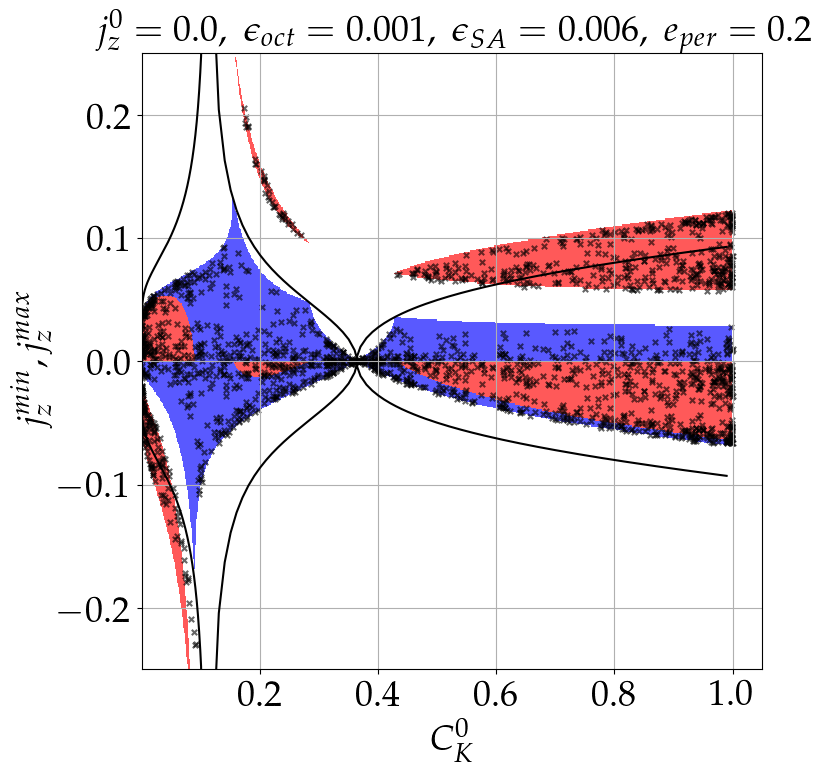}
 \par\end{centering}
\caption{$j^{\text{max}}_z$ and $j^{\text{min}}_z$ vs. $C^0_K$ when $j^0_z=e^0_z=0$ and all other components randomly chosen (uniformly distributed in $j_x$ and $e_y$) for $1000$ realizations of initial conditions. The results of direct numerical integrations of the double averaged equations (up to $\tau=10/\epsilon_\text{oct}$) are shown using black crosses. The envelope of the analytic model of the simple pendulum from \citepaperI is shown using a black curve (depends only on $\epsilon_\text{oct}$). Blue regions denote the analytic prediction for rotating pendulums and red regions denote the analytic prediction for librating pendulums through Equations \protect\ref{eq:librating_possible_from_jz0_0}-\protect\ref{eq:rotation_envelope_negative_fOmega_jz0_0_no_libration}.}
 \label{fig:jz_minmax_vs_CK_numerical_all_points}
\end{figure}

Within the approximation of constant $C_K$, one way of obtaining the value of maximal $j^0_z$ allowing a flip is evaluating the maximal deviation of $j_z$ from $j^0_z=0$. For example, starting with $j^0_z=0$ we have for the maximal available value of $j^\text{max}_z$: Given $\epsilon_\text{oct},\epsilon_\text{SA}$ and $e_{\text{per}}$ and $C^0_K$ the condition for a possibility of libration of the pendulum is 
 \begin{equation}
 \frac{\Delta^2}{2\epsilon_\text{oct}\left| \left\langle f_{j} \right\rangle \left\langle f_{\Omega} \right\rangle \right|}\leq2.\label{eq:librating_possible_from_jz0_0}  
 \end{equation}
 If $\left\langle f_{\Omega} \right\rangle > 0$ (i.e, $C^0_K\gtrsim 0.112$) and libration is possible (Equation \ref{eq:librating_possible_from_jz0_0}) the envelopes of the different regimes are given by 
 \begin{equation}
   \begin{aligned}
     0 \leq j^\text{max, rotation}_z&\leq\frac{\Delta}{\left\langle f_{\Omega} \right\rangle}\\
     2\frac{\Delta}{\left\langle f_{\Omega} \right\rangle} \leq j^\text{max, libration}_z& \leq \frac{\Delta}{\left\langle f_{\Omega} \right\rangle} + 2\sqrt{\epsilon_\text{oct}\left|\frac{\left\langle f_{j} \right\rangle}{\left\langle f_{\Omega} \right\rangle}\right|}.
   \end{aligned}
 \end{equation}
 If libration is not possible, the envelope is given by 
 \begin{equation}
   0 \leq j^\text{max, rotation}_z \leq \frac{\Delta-\sqrt{\Delta^2 - 4\epsilon_\text{oct}\left|\left\langle f_{j} \right\rangle \left\langle f_{\Omega} \right\rangle\right|}}{\left\langle f_{\Omega} \right\rangle}.
 \end{equation}
 If $\left\langle f_{\Omega} \right\rangle < 0$ (i.e, $C^0_K\lesssim 0.112$) and libration is possible the envelopes of the different regimes are given by 
 \begin{equation}
   \begin{aligned}
     \frac{\Delta}{\left\langle f_{\Omega} \right\rangle} + 2\sqrt{\epsilon_\text{oct}\left|\frac{\left\langle f_{j} \right\rangle}{\left\langle f_{\Omega} \right\rangle}\right|} \leq j^\text{max, rotation}_z&\leq\frac{\Delta-\sqrt{\Delta^2 + 4\epsilon_\text{oct}\left|\left\langle f_{j} \right\rangle \left\langle f_{\Omega} \right\rangle\right|}}{\left\langle f_{\Omega} \right\rangle}.\\
     0 \leq j^\text{max, libration}_z& \leq \frac{\Delta}{\left\langle f_{\Omega} \right\rangle} + 2\sqrt{\epsilon_\text{oct}\left|\frac{\left\langle f_{j} \right\rangle}{\left\langle f_{\Omega} \right\rangle}\right|}.
   \end{aligned}
 \end{equation}
 If libration is not possible, the envelope is given by 
 \begin{equation}
   0 \leq j^\text{max, rotation}_z \leq \frac{\Delta-\sqrt{\Delta^2 + 4\epsilon_\text{oct}\left|\left\langle f_{j} \right\rangle \left\langle f_{\Omega} \right\rangle\right|}}{\left\langle f_{\Omega} \right\rangle}.\label{eq:rotation_envelope_negative_fOmega_jz0_0_no_libration}
 \end{equation}
 Similar expressions can be obtained for $j^\text{min}_z$. The maximal and minimal values of $j_z$ obtained in $1000$ numerical simulations with $j^0_z=e^0_z=0$ (i.e $i^0=0,\omega^0=0$) and randomly chosen initial conditions (uniformly distributed in $j_x$ and $e_y$) are plotted using black crosses in Figure \ref{fig:jz_minmax_vs_CK_numerical_all_points} for $\epsilon_{\text{oct}}=0.001,\epsilon_{\text{SA}}=0.006$ and $e_{\text{per}}=0.2$. The black line in Figure \ref{fig:jz_minmax_vs_CK_numerical_all_points} comes from Equation 12 of \citepaperI and is the envelope of both $j^{\text{max}}_z$ and $j^{\text{min}}_z$ in the analytic model when Brown's Hamiltonian is not taken into account (i.e $\epsilon_{\text{SA}}=0$). The red regions denote allowed results from librating pendulums while blue regions denote rotating pendulums (through Equations \ref{eq:librating_possible_from_jz0_0}-\ref{eq:rotation_envelope_negative_fOmega_jz0_0_no_libration}). As can be seen, the shaded regions agree with the envelope of the numerically available $j^\text{max}_z$ and $j^{\text{min}}_z$ and predict the splitting of allowed regions between librating and rotating pendulums. The analytic model allows for a prediction of the maximal deviation for each set of initial conditions and these are shown in a similar plot in appendix \ref{app:jz_minmax_vs_CK_numerical_and_analytic_all_points}.

\subsubsection{Dependence on $\epsilon_\text{oct},\epsilon_\text{SA}$ and $e_{\text{per}}$}\label{subsubsubsec:dependence}
\begin{figure}
 \begin{centering}
 \includegraphics[scale=0.21]{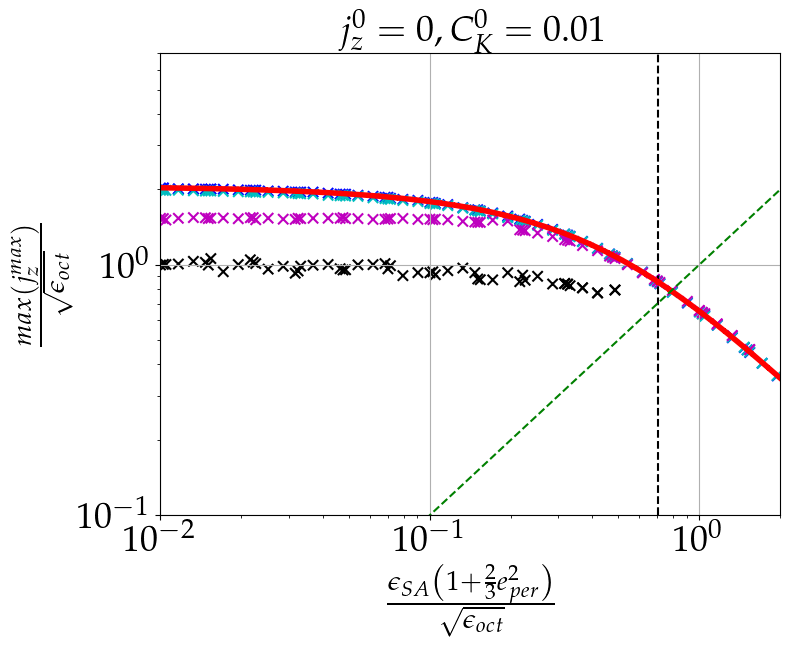}
 \includegraphics[scale=0.21]{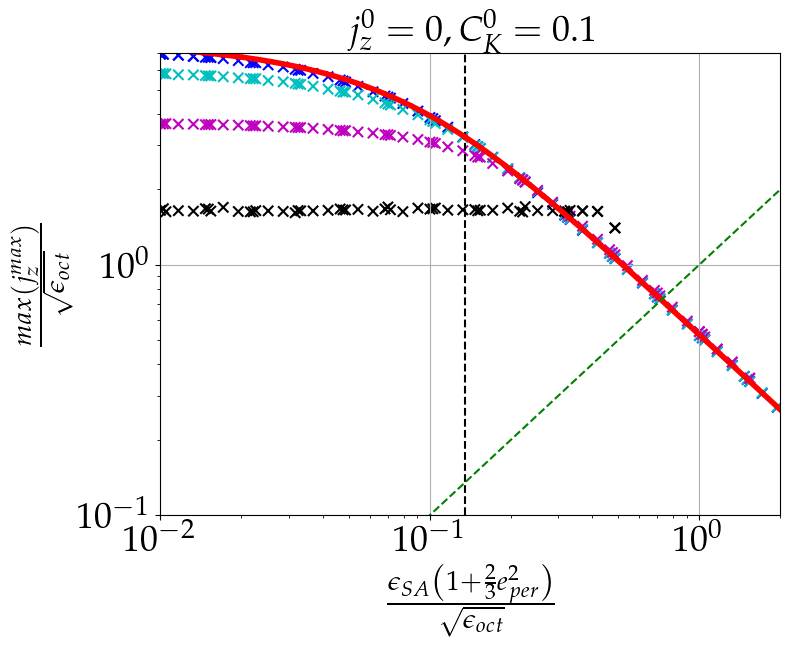}
 \par\end{centering}
 \begin{centering}
 \includegraphics[scale=0.21]{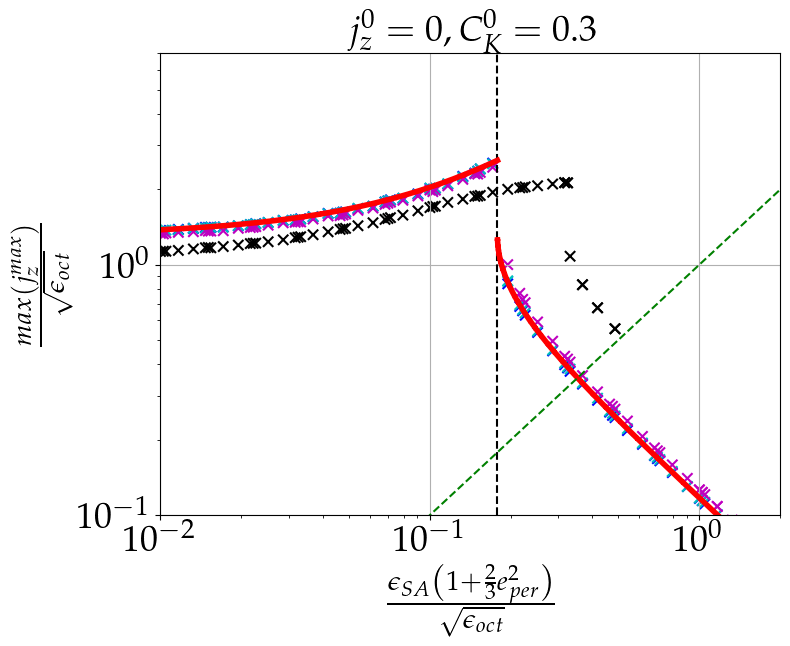}
 \includegraphics[scale=0.21]{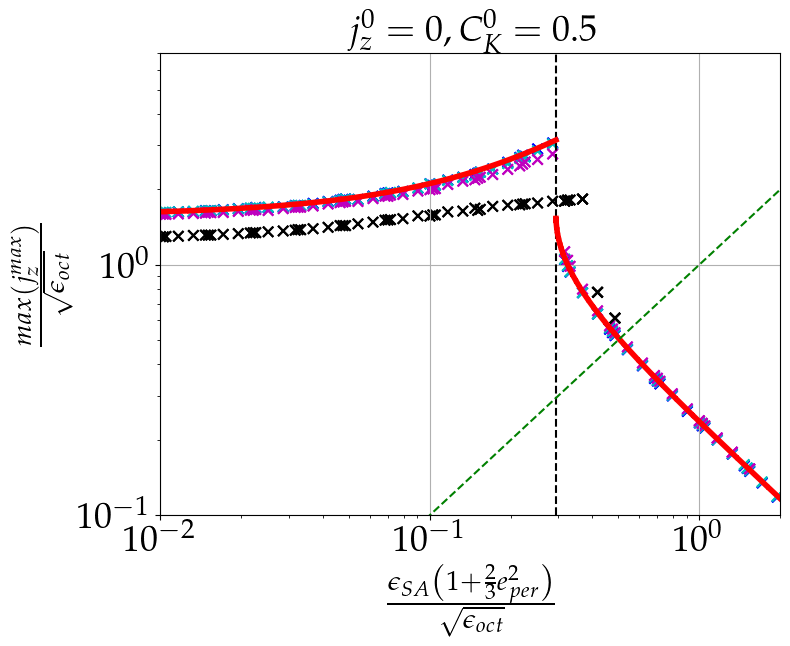}
 \par\end{centering}
 \begin{centering}
 \includegraphics[scale=0.21]{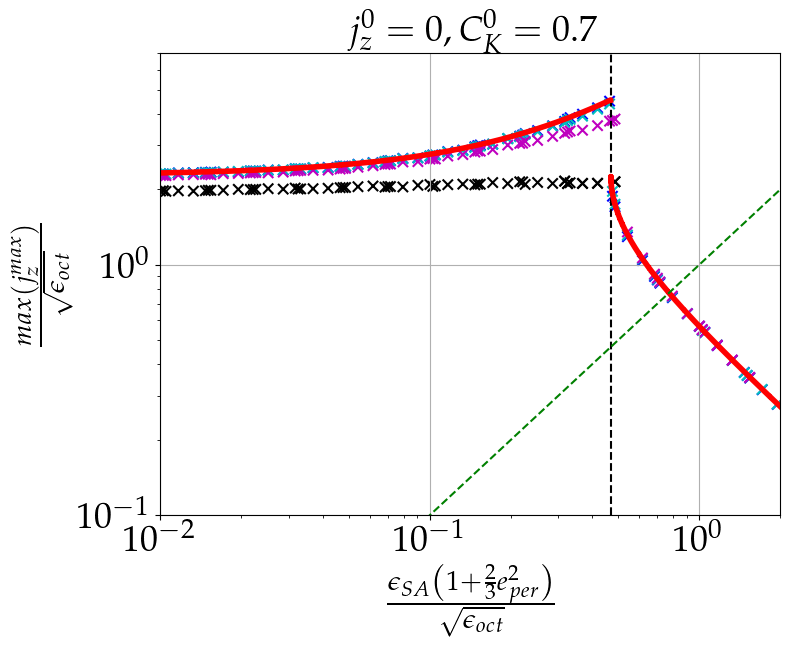}
 \includegraphics[scale=0.21]{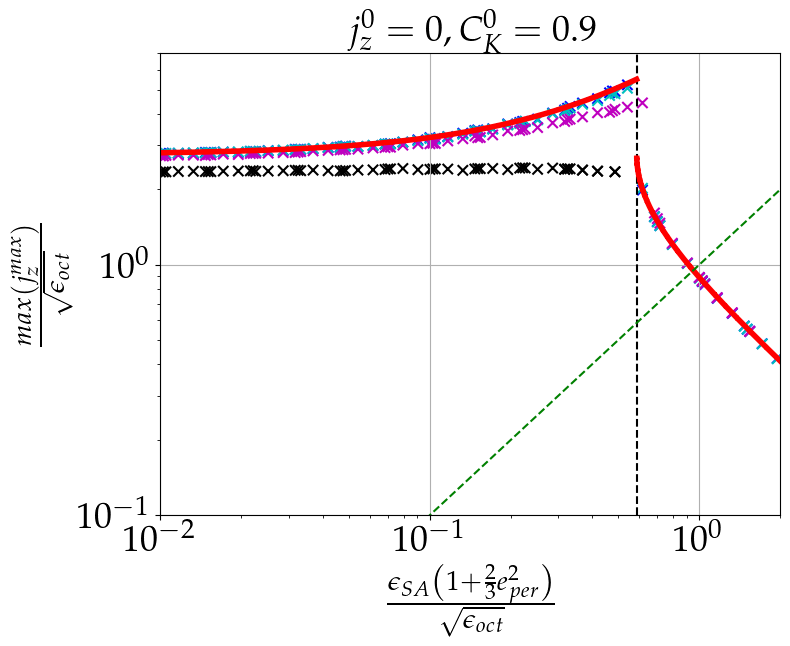}
 \par\end{centering}
 % \begin{centering}
 % \includegraphics[scale=0.19]{jzmax_vs_scaling_numerical_colored_pendulum_red_Ck0_0.5.png}
 % \includegraphics[scale=0.19]{jzmax_vs_scaling_numerical_colored_pendulum_red_Ck0_0.8.png}
 % \par\end{centering}
 %  \begin{centering}
 % \includegraphics[scale=0.19]{jzmax_vs_scaling_numerical_colored_pendulum_red_Ck0_0.9.png}
 % \includegraphics[scale=0.19]{jzmax_vs_scaling_numerical_colored_pendulum_red_Ck0_0.99.png}
 % \par\end{centering}
 \caption{The maximal attainable value of $\frac{j^\text{max}_z}{\sqrt{\epsilon_\text{oct}}}$ from $j^0_z=e^0_z=0$ vs. the scaling factor $s$ (Equation \ref{eq:scaling_factor}). Each panel has a \textbf{different} value of the initial $C_K$ shown above the plot. The vertical black dashed curve denotes the value of $s^\text{libration \ possible}$ (Equation \ref{eq:scaling_factor_librating_possible}). Each cross is a result of $36$ numerical integrations with the same $\epsilon_\text{oct},\epsilon_\text{SA},e_{\text{per}},\omega^0=0,j^0_z=0,C^0_K$ and denote the maximal value of $j^\text{max}_z$ over a scan of all possible values of $\Omega^0$ (equally spaced by $10^\circ$ between $0$ and $360^\circ$). Shown are four values of $\epsilon_\text{oct}=0.0001$ (blue), $0.001$ (cyan), $0.01$ (magenta) and $0.1$ (black). The values of $\epsilon_\text{SA}$ are scanned using a logarithmic spacing of $19$ values between $0.0001$ and $0.1$. The values of $e_{\text{per}}$ are taken from the set $0.05, 0.2, 0.5, 0.7$ and $0.9$. The red curve denotes the prediction of the simple pendulum model using Equations \ref{eq:librating_possible_from_jz0_0}-\ref{eq:rotation_envelope_negative_fOmega_jz0_0_no_libration}. The green dashed line is a line of $y=x$ schematically presenting the scale of the fluctuations effect from \citet{haim18}.\label{fig:jzmaxmax_vs_scaling_factor}}
\end{figure}

Using the scaling factor defined in section \ref{subsec:scaling} an evaluation of the maximal available deviation of $j_z$ from $j^0_z=0$ can be presented. For each $C^0_K$ (and $j^0_z=0$) there is a special value of $s$ separating the parameter space of $\epsilon_\text{oct},\epsilon_\text{SA}$ and $e_{\text{per}}$ between pendulums that can librate (for some $\Omega^0_e$) and pendulums that can only rotate:
\begin{equation}
  s^\text{libration \ possible}<\frac{128}{27}\frac{\sqrt{\left|\left\langle f_{j}\right\rangle \left\langle f_{\Omega}\right\rangle \right|}}{\left(\frac{1}{3}+8C_{K}+15\left\langle e_{z}^{2} \right\rangle\right)}\label{eq:scaling_factor_librating_possible}.
\end{equation}
Given the parameters $\epsilon_\text{oct},\epsilon_\text{SA}$ and $e_{\text{per}}$ and initial values $C^0_K$ and $\omega^0=0$ and $j^0_z=0$ the maximal attainable value of $j^\text{max}_z$ can be evaluated both numerically (by scanning all possible values of $\Omega^0$) and analytically using the simple pendulum model (through $\dot{\phi}^\text{max}$ and $\dot{\phi}^\text{min}$). These evaluations are plotted in Figure \ref{fig:jzmaxmax_vs_scaling_factor} showing the agreement between different values of the parameters when scaled using $s$. Each subplot presents results for some value of $C^0_K$ (written above the subplot). We scan 4 values of $\epsilon_\text{oct}=0.0001$ (blue), $0.001$ (cyan), $0.01$ (magenta) and $0.1$ (black). For $\epsilon_\text{SA}$ we scan $19$ values logarithmically spaced between $0.0001$ and $0.1$ and for $e_{\text{per}}$ we use the values $0.05, 0.2, 0.5, 0.7$ and $0.9$. Each cross represents the maximal value of $j^\text{max}_z$ obtained from $36$ simulations with $\Omega_0$ equally spaced by $10^\circ$ between $0$ and $360^\circ$. The red curve denotes the prediction of the simple pendulum model using Equations \ref{eq:librating_possible_from_jz0_0}-\ref{eq:rotation_envelope_negative_fOmega_jz0_0_no_libration}. Note that the deviation of the black crosses (denoting simulations with $\epsilon_\text{oct}=0.1$) from a simple pendulum model is present already in in Figure 4 of \citepaperI presenting the results of a simple pendulum model for the "no Brown's Hamiltonian" equations (i.e when $\epsilon_\text{SA}=0$). As $s$ increases, i.e as $\epsilon_\text{SA}$ increases, a different mechanism, fluctuations in $j_z$ during close approaches \citep{haim18}, becomes dominant as its effect scales as $j^\text{max}_z \sim \epsilon_\text{SA}$ (see Equation 13 of \citet{haim18}). A scale of this effect is schematically plotted in Figure \ref{fig:jzmaxmax_vs_scaling_factor} using a dashed green line. All in all, the inclusion of just Brown's Hamiltonian is important in the portion of phase space of $s$ shown in Figure \ref{fig:jzmaxmax_vs_scaling_factor} for not too low values (where the octupole term is much stronger than Brown's Hamiltonian) and not too high values where other mechanisms must be taken into account as well. As presented in Equations \ref{eq:librating_possible_from_jz0_0}-\ref{eq:rotation_envelope_negative_fOmega_jz0_0_no_libration} and in Figure \ref{fig:jz_minmax_vs_CK_numerical_all_points} there is a difference between the region where $\left\langle f_{\Omega}\right\rangle > 0$ ($C^0_K\gtrsim 0.112$) where the maximal value of $j^\text{max}_z$ is obtained in librating region if they are possible to the region where $\left\langle f_{\Omega}\right\rangle < 0$ ($C^0_K\lesssim 0.112$) where the the maximal value of $j^\text{max}_z$ is obtained from rotating pendulums even when libration is possible. This is the origin of the sharp peaks seen in the panels of Figure \ref{fig:jzmaxmax_vs_scaling_factor} with $C^0_K$ values greater than $\approx0.112$ and are absent for the two upper panels with $C^0_K=0.01,0.1$ with $\left\langle f_{\Omega}\right\rangle < 0$.

\subsection{flip criterion}\label{subsec:flip_criterion}
\begin{figure}
 \begin{centering}
 \includegraphics[scale=0.23]{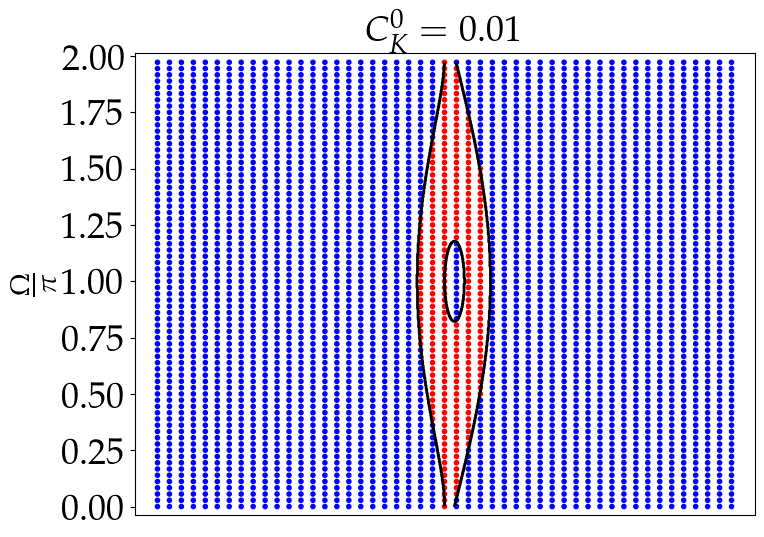}
 \includegraphics[scale=0.24]{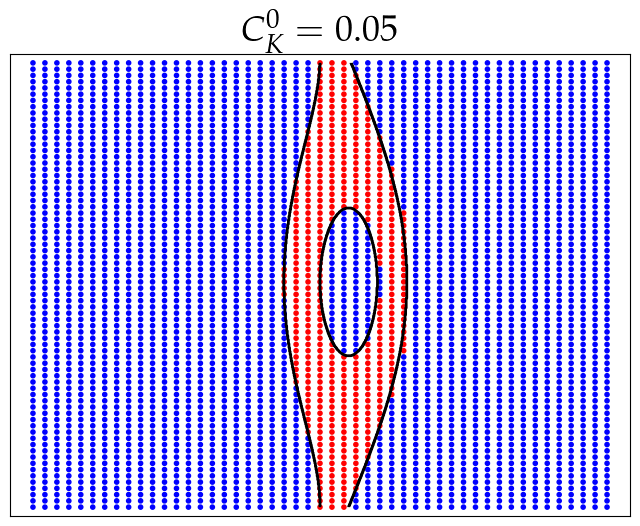}
 \par\end{centering}
 \begin{centering}
 \includegraphics[scale=0.23]{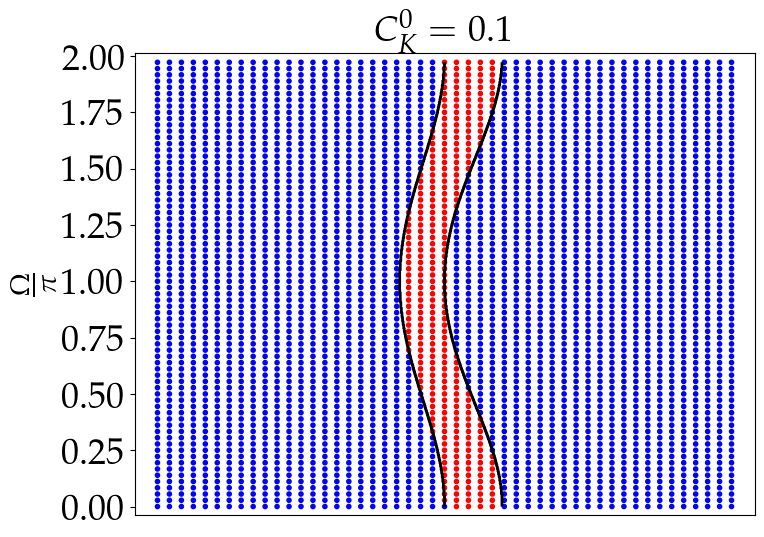}
 \includegraphics[scale=0.24]{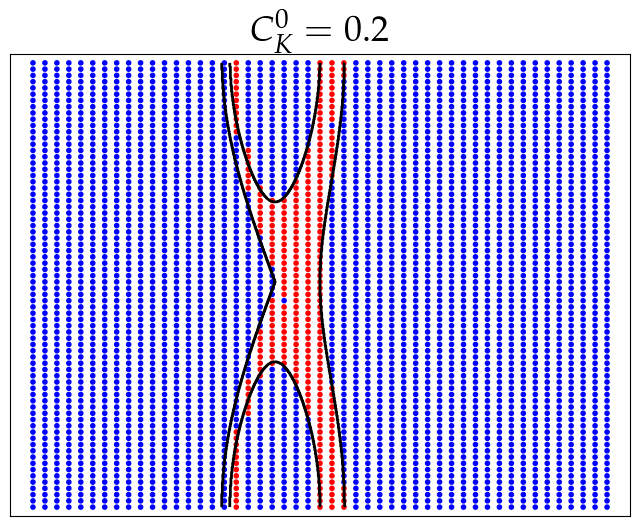}
 \par\end{centering}
 \begin{centering}
 \includegraphics[scale=0.23]{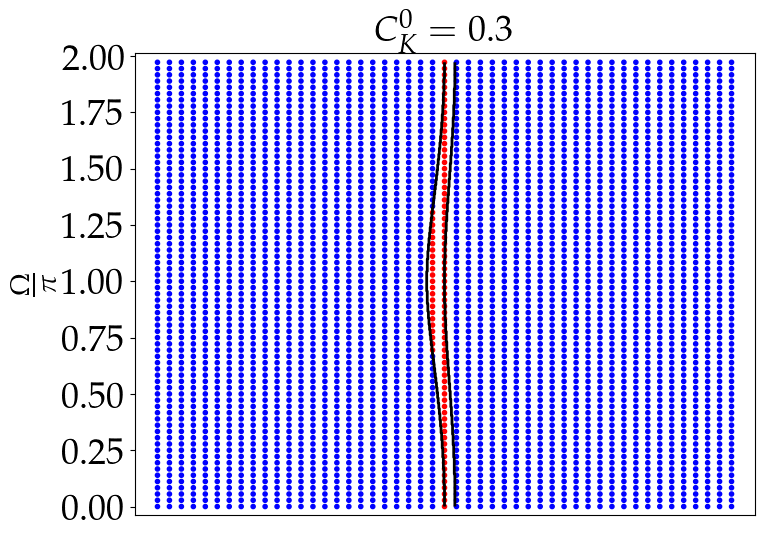}
 \includegraphics[scale=0.24]{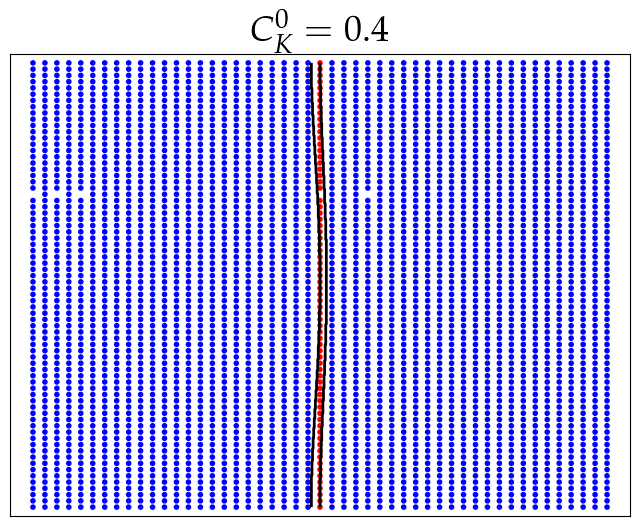}
 \par\end{centering}
 \begin{centering}
 \includegraphics[scale=0.23]{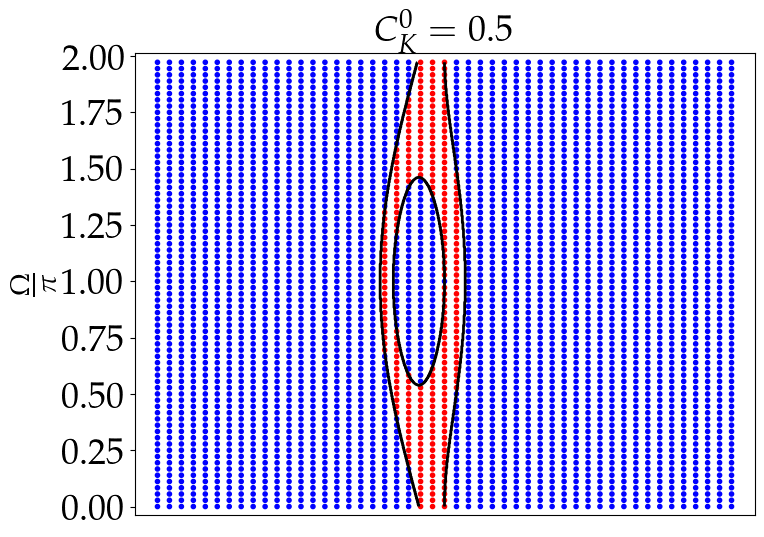}
 \includegraphics[scale=0.24]{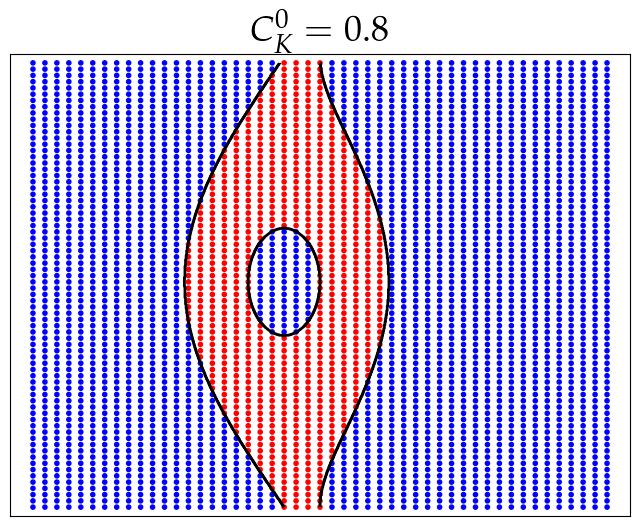}
 \par\end{centering}
  \begin{centering}
 \includegraphics[scale=0.23]{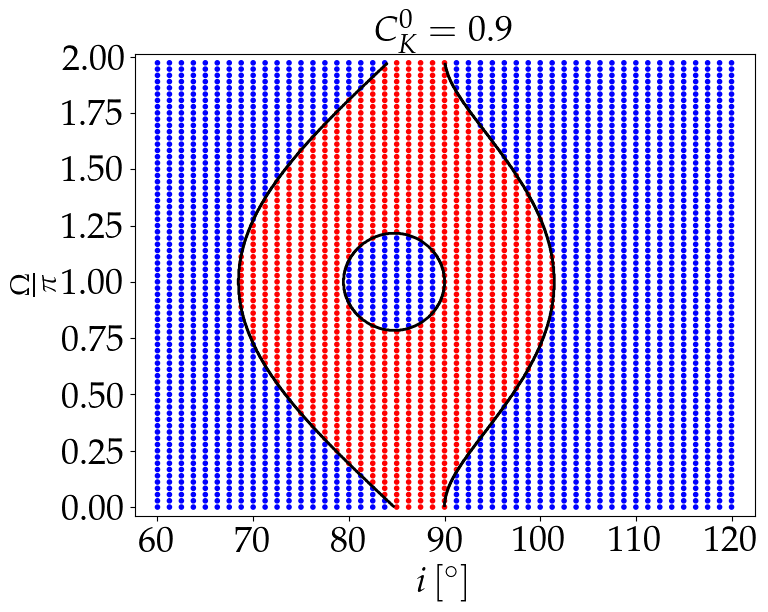}
 \includegraphics[scale=0.24]{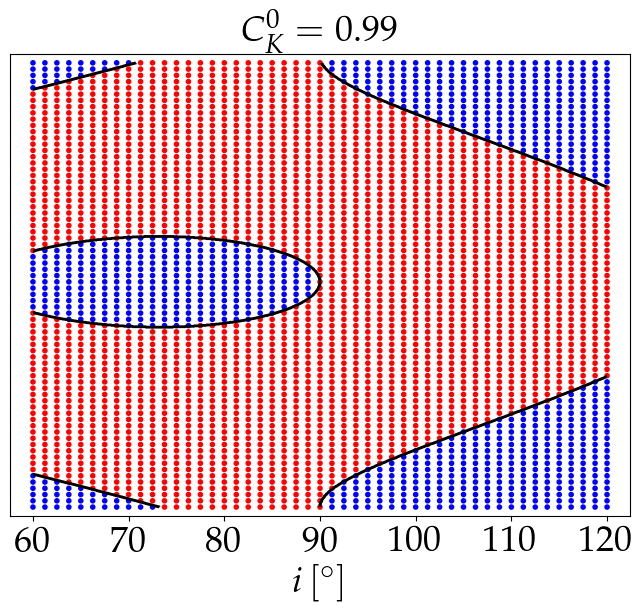}
 \par\end{centering}
 \caption{Parameter space of orbital flips from the solution of the double averaged secular equations (up to $\tau=10/\epsilon_\text{oct}$). All panels have the same values of $\epsilon_{\text{oct}}=0.001,\epsilon_{\text{SA}}=0.006$ and $e_{{\text{per}}}=0.2$. Each panel have a \textbf{different} value of the initial $C_K$ shown above the plot. All integrations have $\omega^0=0$ and the initial values of inclination and $\Omega$ are 
 scanned. Red points denote integrations where a flip occurred and blue points denote these without a flip. Black curves denote the flip criterion of the simple pendulum model (Equations \ref{eq:flip_when_pendulum_librating}- \ref{eq:flip_when_pendulum_rotating} and the separation between librating and rotating pendulum).\label{fig:flipmaps}}
\end{figure}

Another form of a flip criterion is a map from initial conditions to a flip or non flip outcome. Given $\epsilon_{\text{oct}},\epsilon_{\text{SA}}$ and $e_{\text{per}}$ and initial values $j^0_z,C^0_K$ and $\Omega^0_e$ (with $\omega^0=0$), a condition for whether a flip would occur can be obtained using the simple pendulum model. When the pendulum is in libration for example, a flip will occur if 
\begin{equation}
\frac{1}{2}\Delta^2 < E < 2 \epsilon_{\text{oct}} \left| \left\langle f_{j} \right\rangle \left\langle f_{\Omega} \right\rangle \right|  
\end{equation}
where the right hand side is the libration condition and the left hand side is the flip criterion when in libration. When the pendulum is in rotation mode the flip criterion is more complex and is given in appendix \ref{app:explicit_flip_criterion}. A comparison between a numerical \textit{flip map} and the analytic prediction of the simple pendulum model (Equations \ref{eq:flip_when_pendulum_librating}, \ref{eq:flip_when_pendulum_rotating}) as a function of $i^0$ and $\Omega^0$ is shown in Figure \ref{fig:flipmaps} for $\epsilon_{\text{oct}}=0.001,\epsilon_{\text{SA}}=0.006$ and $e_{\text{per}}=0.2$. Each point represent the result of a numerical simulation of the double averaged equations (up to $\tau=10/\epsilon_\text{oct}$) with $\omega^0=0$, and $C^0_K=\left(e^0\right)^2$ written above each subplot. A point is marked red if a flip occurred and blue otherwise. Black curves mark the analytic prediction of the flip criterion using the simple pendulum model via Equations \ref{eq:flip_when_pendulum_librating}, \ref{eq:flip_when_pendulum_rotating}. As can be seen, the black curve follows the border between red and blue points for the wide range of $C^0_K$ values resulting with different forms of a flip map.

\section{Discussion}
In this study, we extended the approach presented in \citepaperI to analytically solve the EKL effect at high eccentricities, incorporating Brown's Hamiltonian. Using the simplified version of Brown's Hamiltonian from \citet{tremaine2023}, we demonstrated that its prominent effect on long-term evolution is an additional precession of the eccentricity vector. The dynamics can be described using a simple pendulum model where the additional term imposes a shift in the the pendulum's velocity. This model allows for the derivation of an explicit flip criterion (Figure \ref{fig:flipmaps}) dependent on whether the equivalent simple pendulum being in rotation or libration (Equations \ref{eq:flip_when_pendulum_librating}, \ref{eq:flip_when_pendulum_rotating}).

The effective equations obtained introduce a scaling factor, $s$ (Equation \ref{eq:scaling_factor}), which measures the importance of including Brown's Hamiltonian relative to the octupole term (Figure \ref{fig:jzmaxmax_vs_scaling_factor}). When the $s\lesssim0.1$ - the dynamics are dominated by the octupole term alone. For $s\gtrsim1$ additional effects need to be considered (e.g, \cite{luo2016,haim18}). This identifies the region of $0.1 \lesssim s \lesssim 1$ where the inclusion of (just) Brown's Hamiltonian is crucial for accurately the high eccentricity dynamics and obtaining a physical understanding of the flip criterion through the analytical model. 

For example, consider the parameters used in this Letter: $\epsilon_\text{oct}=0.001$ and $\epsilon_\text{SA}=0.006$ (with $e_\text{per}=0.2$) leading to a scaling factor $s\approx0.2$. These values correspond to a hierarchical system with $m_\text{per}/m\sim300$ and $a_\text{per}/a\sim200$. Consider a Jupiter-sized planet orbiting an M-dwarf of mass $m_\text{per}=0.3M_\odot$ on an orbit with $a_\text{per}=5\text{AU}$. A satellite orbiting the Jupiter at $a\sim0.025\text{AU}$ has a scaling factor $s\approx0.2$, indicating that including (just) Brown's Hamiltonian in addition to the octupole term is necessary for accurately describing its dynamics. In the high eccentricity regime, this system can be analytically investigated using the simple pendulum model.

% In conclusion, our study provides a comprehensive analytical framework for understanding the long-term evolution of high eccentricity orbits in hierarchical three-body systems, emphasizing the critical role of Brown's Hamiltonian. This approach enhances our ability to predict and analyze the dynamical behavior of various astrophysical systems, such as exoplanets and satellite systems, using a simplified yet accurate model.

% \section*{Acknowledgements}

% We thank ...

%%%%%%%%%%%%%%%%%%%%%%%%%%%%%%%%%%%%%%%%%%%%%%%%%%
\section*{Data Availability}

The codes used in this article will be shared on reasonable request.



\bibliographystyle{mnras}
\bibliography{HB3_is_a_pendulum} % if your bibtex file is called example.bib

%\begin{thebibliography}{99}
%\end{thebibliography}

%%%%%%%%%%%%%%%%%%%%%%%%%%%%%%%%%%%%%%%%%%%%%%%%%%

%%%%%%%%%%%%%%%%% APPENDICES %%%%%%%%%%%%%%%%%%%%%

\appendix

\section{Explicit flip criterion}\label{app:explicit_flip_criterion}
A flip - zero crossing of $j_z$ - happens when the product $j^{\text{max}}_zj^{\text{min}}_z < 0$. From Equation \ref{eq:phidot} we obtain expressions for $j^{\text{max}}_z$ and $j^{\text{min}}_z$: For $C_{K}^{0}\gtrsim 0.112$ where $\left\langle f_\Omega \right\rangle > 0$ 
\begin{equation}
\begin{aligned}
  j^{\text{max}}_z &= \frac{1}{\left\langle f_\Omega \right\rangle}\left(\dot{\phi}^{\text{max}} +\Delta \right)\\
  j^{\text{min}}_z &= \frac{1}{\left\langle f_\Omega \right\rangle}\left(\dot{\phi}^{\text{min}} +\Delta \right)
\end{aligned}  
\end{equation}
and vice-versa where $\left\langle f_\Omega \right\rangle < 0$.
The condition for a flip depends on the initial conditions and specifically differs between two separate states of the pendulum: (i) Libration: When $E < 2 \left| \epsilon_{\text{oct}} \left\langle f_{j} \right\rangle \left\langle f_{\Omega} \right\rangle \right|$ the pendulum is in libration mode, i.e $\dot{\phi}^{\text{max}}=\sqrt{2E}$ and $\dot{\phi}^{\text{min}}=-\sqrt{2E}$.
% \dot{\phi}$ crosses zero and
% \begin{equation}
%   \dot{\phi}^{\text{max,min}}=\pm \sqrt{2E}.
% \end{equation}
In this case, $j_z^{\text{max}}$ and $j_z^{\text{min}}$ have different signs if 
\begin{equation}
\Delta < \sqrt{2E}  
\end{equation}
and this is the flip criterion if the pendulum is in libration. Given $\epsilon_{\text{oct}},\epsilon_{\text{SA}}$ and $e_{\text{per}}$ and initial values $j^0_z,C^0_K$ and $\Omega^0_e$ if both inequalities are obeyed 
\begin{align}
  \left(\left\langle f_{\Omega}\right\rangle j_{z}^{0}\right)^{2}-2\left\langle f_{\Omega}\right\rangle \Delta j_{z}^{0}+2\left|\epsilon_{\text{oct}}\left\langle f_{j}\right\rangle \left\langle f_{\Omega}\right\rangle \right|\left(1-\cos\phi^{0}\right)&>0\cr
  \frac{1}{2}\left(\left\langle f_{\Omega}\right\rangle j_{z}^{0}-\Delta\right)^{2}-\left|\epsilon_{\text{oct}}\left\langle f_{j}\right\rangle \left\langle f_{\Omega}\right\rangle \right|\left(1+\cos\phi^{0}\right)&<0\label{eq:flip_when_pendulum_librating}
\end{align}
a flip occurs. (ii) Rotation: When $E > 2 \left| \epsilon_{\text{oct}} \left\langle f_{j} \right\rangle \left\langle f_{\Omega} \right\rangle \right|$ the pendulum is in rotation mode, and $\dot{\phi}$ does not change its sign. Since $\Delta>0$ there is no flip if $\dot{\phi}^0>0$. If $\dot{\phi}^0<0$, $\dot{\phi}^{\text{max}}=-\sqrt{2\left(E-2\left|\epsilon_{\text{oct}}\left\langle f_{j}\right\rangle \left\langle f_{\Omega}\right\rangle \right|\right)}$ and $\dot{\phi}^{\text{min}}=-\sqrt{2E}$. A flip occurs if
\begin{equation}
  \left|\dot{\phi}^{\text{max}}\right|<\Delta<\left|\dot{\phi}^{\text{min}}\right|
\end{equation}
i.e if the following inequalities are obeyed 
\begin{align}
  \left(\left\langle f_{\Omega}\right\rangle j_{z}^{0}\right)^{2}-2 \Delta\left\langle f_{\Omega}\right\rangle j_{z}^{0}+2\left|\epsilon_{\text{oct}}\left\langle f_{j}\right\rangle \left\langle f_{\Omega}\right\rangle \right|\left(1-\cos\phi^{0}\right)&>0\cr
  \left(\left\langle f_{\Omega}\right\rangle j_{z}^{0}\right)^{2}-2\Delta\left\langle f_{\Omega}\right\rangle j_{z}^{0}-2\left|\epsilon_{\text{oct}}\left\langle f_{j}\right\rangle \left\langle f_{\Omega}\right\rangle \right|\left(1+\cos\phi^{0}\right)&<0\cr
  \frac{1}{2}\left(\left\langle f_{\Omega}\right\rangle j_{z}^{0}-\Delta\right)^{2}-\left|\epsilon_{\text{oct}}\left\langle f_{j}\right\rangle \left\langle f_{\Omega}\right\rangle \right|\left(1+\cos\phi^{0}\right)&>0\cr
  \left\langle f_{\Omega}\right\rangle j^0_z-\Delta&<0\label{eq:flip_when_pendulum_rotating}
\end{align}

The third inequality is the rotation condition. The fourth inequality is the negative $\dot{\phi}^0$ condition and the first two inequalities are the flip criterion for the rotation region.

% \section{$\MakeLowercase{j}^{\MakeLowercase{max}}_\MakeLowercase{z}$ and $\MakeLowercase{j}^{\MakeLowercase{min}}_\MakeLowercase{z}$ from $\MakeLowercase{j}^0_\MakeLowercase{z}=\MakeLowercase{e}^0_\MakeLowercase{z}=0$ }\label{app:jz_minmax_vs_CK_numerical_and_analytic_all_points}
\section{$j^{\text{max}}_z$ and $j^{\text{min}}_z$ from $j^0_z=e^0_z=0$ }\label{app:jz_minmax_vs_CK_numerical_and_analytic_all_points}
\begin{figure}
 \begin{centering}
 \includegraphics[scale=0.45]{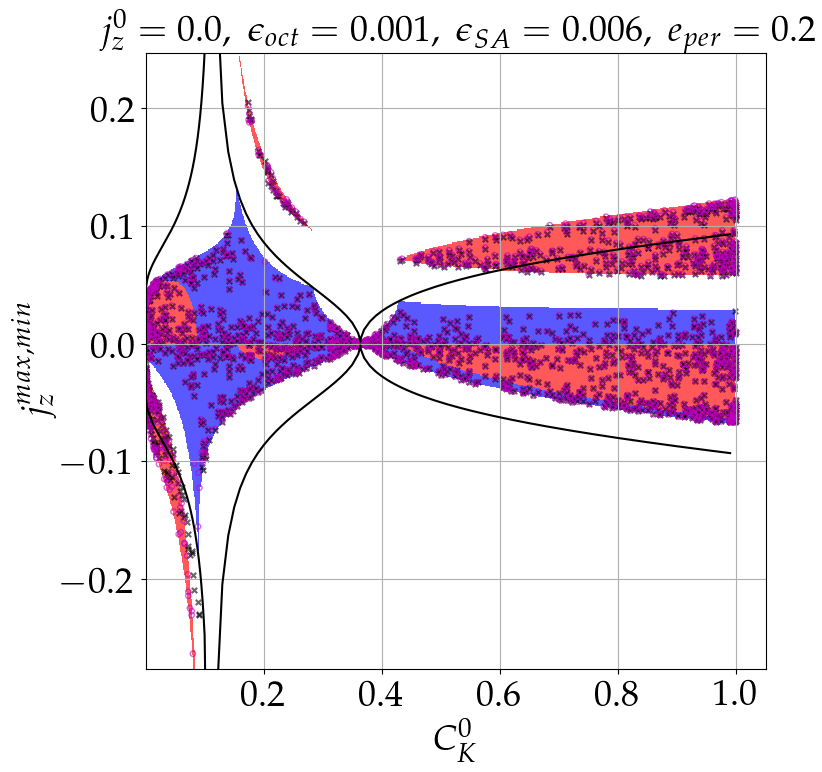}
 \par\end{centering}
 \caption{$j^{\text{max}}_z$ and $j^{\text{min}}_z$ vs. $C^0_K$ when $j^0_z=e^0_z=0$ and all other components randomly chosen (uniformly distributed in $j_x$ and $e_y$). The results of direct numerical integrations of the double averaged equations (up to $\tau=10/\epsilon_\text{oct}$) are shown using black crosses. The envelope of the analytic model of the simple pendulum from \citepaperI is shown using a black curve (depends only on $\epsilon_\text{oct}$). Blue regions denote the analytic prediction for rotating pendulums and red regions denote the analytic prediction for librating pendulums. Magenta open circles denote the prediction of the simple pendulum model for each set of initial conditions.}\label{fig:jz_minmax_vs_CK_numerical_and_analyticall_points}
\end{figure}
 In Figure \ref{fig:jz_minmax_vs_CK_numerical_and_analyticall_points} we plot the same information as Figure \ref{fig:jz_minmax_vs_CK_numerical_all_points} but we add the prediction of the simple pendulum model for each set of random initial conditions using magenta open circles. As can be seen there, the predictions of Equations \ref{eq:jzmax} for different values of $\Omega^0_e$ agree with the numerical results to an excellent approximation.

%%%%%%%%%%%%%%%%%%%%%%%%%%%%%%%%%%%%%%%%%%%%%%%%%%

% Don't change these lines
\bsp	% typesetting comment
\label{lastpage}
\end{document}